\documentclass[prl,twocolumn,preprintnumbers,amsmath,amssymb]{revtex4}

\usepackage{graphicx}% Include figure files
\usepackage{dcolumn}% Align table columns on decimal point
\usepackage{bm}% bold math
\usepackage{color}
\usepackage{amsmath}
\usepackage[T1]{fontenc}
\usepackage[utf8]{inputenc}
\usepackage{lmodern}
\usepackage[version=3]{mhchem}
\usepackage{array}

\begin{document}

%\title{Direct evidence of nanoscale electrochemical wear on atomically smooth diamond carbon thin films at high sliding speeds}

\title{Electrostatically tunable adhesion in a high speed sliding interface}

\newcommand {\HGST}{Western Digital Company, Recording Sub System Staging and Research, San Jose, CA 95135 USA.}

\author{Sukumar Rajauria}
\altaffiliation{email: sukumar.rajauria@wdc.com}
\affiliation{\HGST}
\author{Oscar Ruiz}
\affiliation{\HGST}
\author{Sripathi V. Canchi}
\affiliation{\HGST}
\author{Erhard Schreck}
\affiliation{\HGST}
\author{Qing Dai}
\affiliation{\HGST}

\begin{abstract}
Contact hysteresis between sliding interfaces is a widely observed phenomenon from macro- to nano- scale sliding interfaces. Most of such studies are done using an atomic force microscope (AFM) where the sliding speed is a few $\mu m /s$. Here, we present a unique study on stiction between the head-disk interface of commercially available hard disk drives, wherein vertical clearance between the head and the disk is of the same order as in various AFM based fundamental studies, but with a sliding speed that is nearly six orders of magnitude higher. We demonstrate that  although the electrostatic force (DC or AC voltage) is an attractive force, the AC voltage induced out-of-plane oscillation of the head with respect to disk is able to suppress completely the contact hysteresis.
\end{abstract}

\date{\today}
\maketitle

%pacs{}
%\keywords{}


\maketitle
%******************************************************************************************************
The ability to reduce stiction in sliding surfaces is extremely important for a variety of applications ranging from very-high-density data storage ($>$ 1 Tb/inch$^{2}$) \cite{Bhushan, WoodIEEE00, PantaziIBM08} to nanoscale manufacturing in micro or nano mechanical systems (M/NEMS) \cite{TsengIBM08, DienwiebelPRL04, KushmerickTL01}. Our research focuses on data storage in hard disk drives, in which the recording heads fly at sub-nanometer spacing from the disk and move at relative sliding speeds of 5-40 $m/s$. Maintaining good tribological properties such as low wear and stiction of such a high speed sliding interface at low clearance is critical for the long term reliability of hard disk drives \cite{MarchonAT03}. Any contact between the head and the disk leads to friction and wear of the head overcoat layer, adversely impacting the disk drive performance \cite{suhTL06, ChenTL14, RajauriaAPL15, RajauriaSR16}. 

A unique way of reducing the friction during contact is through externally imposed oscillations of small amplitude and energy. Previous experimental and theoretical work to reduce the friction at a sliding interface explored the use of external excitation with either surface acoustic waves or electrostatic forces to modulate the out-of-plane or in-plane motion \cite{FridmanJAP59, SocoliucScience06, LantzNatureNano09, PedrazACS15}. The out-of-plane vibrations reduce the friction by momentarily separating the interfering surfaces such that the friction is zero during this time. Consequently, the time averaged friction is reduced below the value that occurs in the absence of oscillation. In this letter, we describe an experimental study of contact events at the head-disk interface, focusing on hysteresis in contact cycle. We show that although the electrostatic force (DC or AC voltage) at the interface is an attractive force, the AC voltage induced oscillation can significantly reduce the contact hysteresis. We provide a fully quantitative analysis of the hysteresis in such a high sliding speed system which shows that the AC voltage induced oscillation has a dominant influence on suppressing the contact hysteresis.

%******************************************************************************************************
\begin{figure}[htbp]
\begin{center}
\includegraphics[width=3.6in]{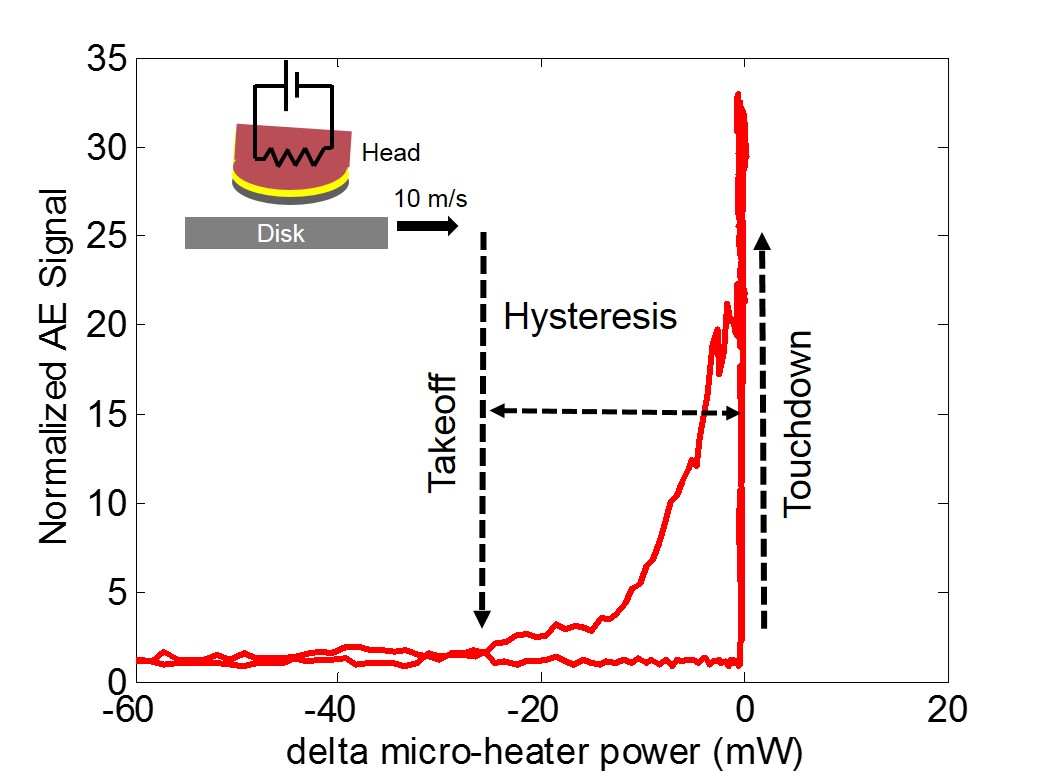}
\end{center}
\caption {Main: Hysteresis touchdown behavior: Normalized acoustic emission signal as a function of delta micro-heater power.  Inset shows the schematic of high sliding speed head-disk interface. }
    \label{fig:1}
\end{figure}
%******************************************************************************************************

%******************************************************************************************************
\begin{figure*}[htbp]
\begin{center}
\includegraphics[width=1\textwidth]{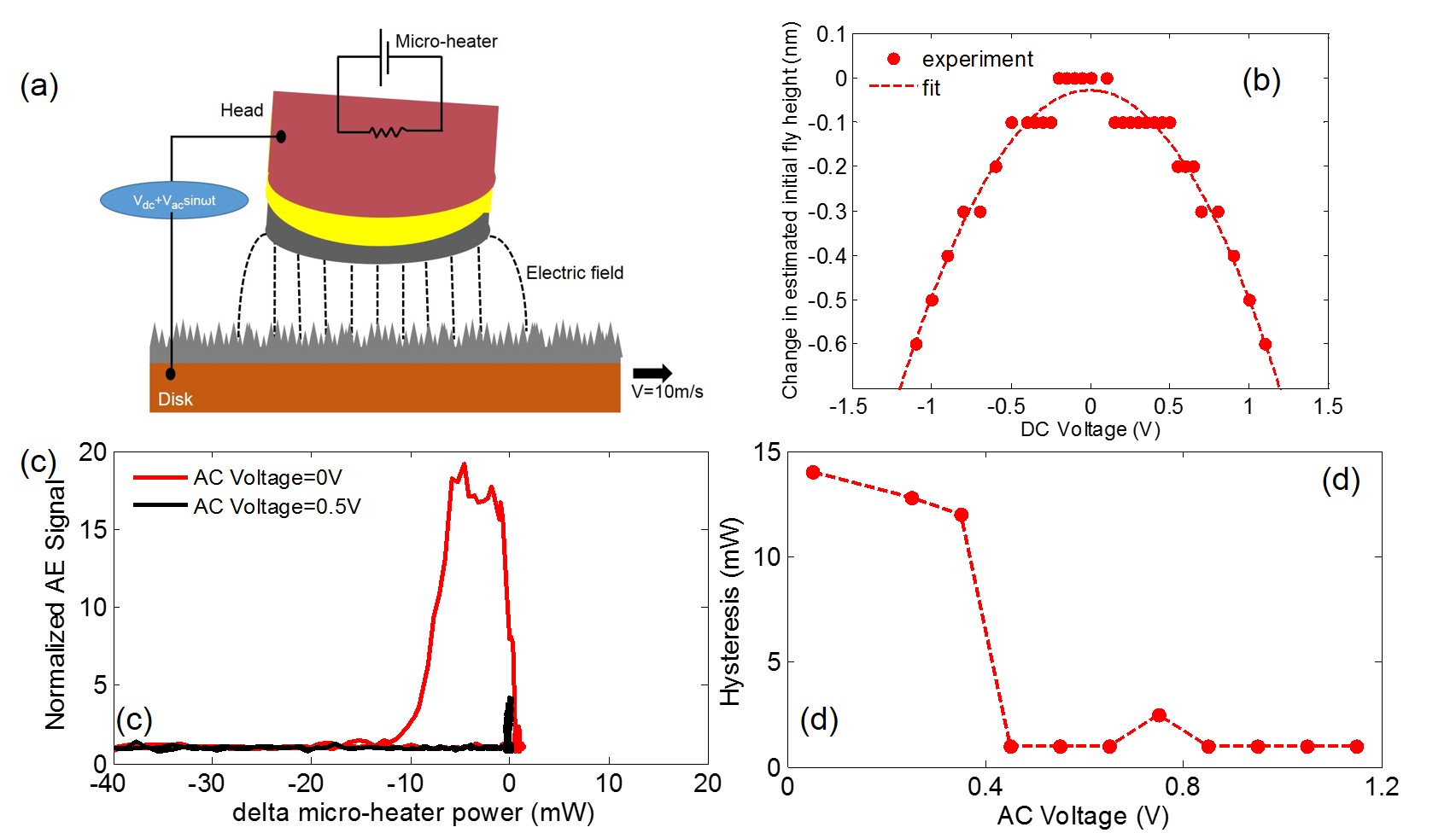}
\end{center}
\caption {Experiment: (a) Schematic cartoon of the head disk interface inside commercial hard disk drives. (d) Shows the reduction in the estimated fly-height as a function of DC voltage. (c) Contact cycle with and without vertical oscillation: Normalized acoustic emission signal as a function of delta micro-heater power. (d) Touchdown/Takeoff hysteresis as a function of AC voltage at the head-disk interface at an oscillating frequency of 10.5 $kHz$.}
    \label{fig:2}
\end{figure*}
%******************************************************************************************************

The disk is fabricated by depositing a magnetic multilayer structure onto a glass substrate, coating it with $\approx$ 3 $nm$ of amorphous nitrogenated carbon (protective overcoat layer), and finally covering it with a molecular layer of perfluoropolyether lubricant ($\approx$ 1 $nm$ thick). The head is coated with 1.4 $nm$ of diamond like carbon on the air-bearing surface (ABS). The typical roughness $R_a$ of the disk and the head surfaces are both 0.4 $nm$. Thus, the head disk interface is a smooth carbon-lubricant-carbon high speed sliding interface.  The head surface is carefully etched such that self pressurized air lift keeps it passively afloat at a fixed clearance over the disk \cite{cml, ZhengTL10}. The initial nominal  clearance (physical gap) between the head and the disk is typically 10 $nm$. Further, clearance is controlled using an embedded micro-heater in the head (see Figure 1 inset). The micro-heater generates a localized protrusion on the head surface, thus bringing it in contact with the disk \cite{ZengIEEE11}. Contact between the head and the disk is detected using an acoustic emission (AE) sensor, which detects elastic propagating waves generated during the head-disk contact events \cite{BhushanIEEE03}.

%******************************************************************************************************
\begin{figure*}[htbp]
\begin{center}
\includegraphics[width=1\textwidth]{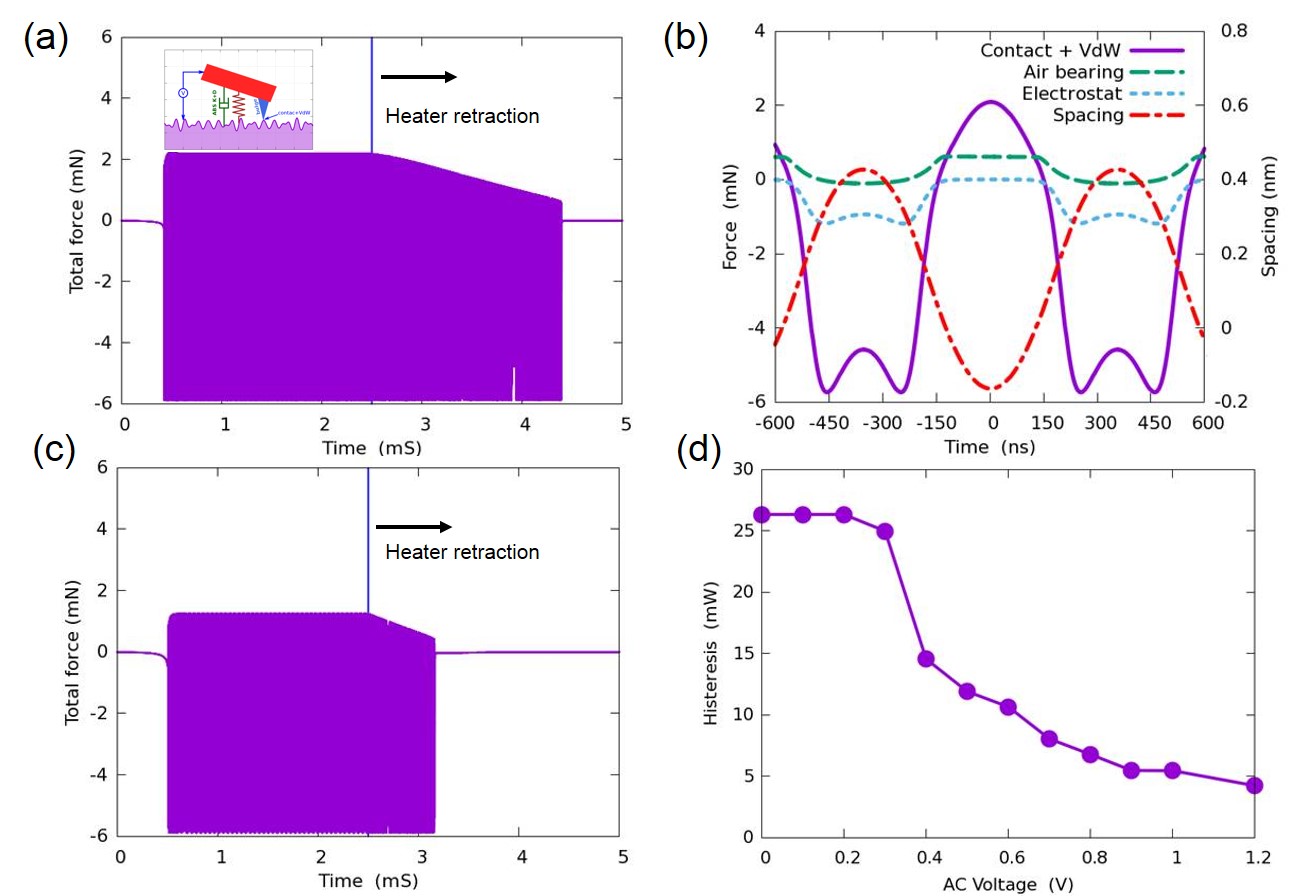}
\end{center}
\caption {Simulation: (a), (c) Shows the complete time domain response of the force at the head-disk interface without (AC Voltage =0$V$, f=10.5$kHz$) and with (AC Voltage =0.6 $V$, f=10.5 $kHz$) AC induced oscillation respectively. (b) Shows the zoom of total force (Purple from Fig3(c)), electrostatic (Blue) and air bearing forces (Green) along with spacing history (Red) for a short time window. (d) Shows the contact hysteresis as a function of the AC voltage at a fixed oscillating frequency of 10.5 $kHz$.}
    \label{fig:3}
\end{figure*}
%******************************************************************************************************

Figure 1 shows a typical contact cycle between the head and the rotating disk. As the micro-heater power increases, a protrusion develops on the head (lateral dimensions $\approx$ 5 $\mu m$ $\times$ 2 $\mu m$ and radius of curvature of around 20 $mm$), bringing it closer to the disk. At a certain micro-heater power (known as touchdown power), contact with the disk is detected as a sharp increase in the AE signal. Afterwards, the micro-heater power is reduced to separate the head from the disk. For the take-off, the heater power needs to be reduced considerably compare to the touchdown power to separate the contacting surfaces, leading to the adhesive hysteresis between the touchdown and take-off condition.

A unique way of controlling and ultimately manipulating the stiction during contact is through externally imposed oscillations of small amplitude and energy. In order to study the effect of vertical modulation on stiction, we applied a low frequency oscillating voltage between the head and disk. The applied oscillating frequency is much lower than the mechanical ABS modes of the head structure (see supplementary). The AC voltage at the interface leads to a small amplitude oscillation of the head described by: $z=z_{o}.sin(2\omega t)$ \cite{KniggeIEEE04}. With voltage oscillation turned-on, the micro-heater power is increased until the AE signal increases sharply. Figure 2(c) shows a contact cycle between the head and the disk with and without oscillation. Clearly, the presence of oscillation significantly suppresses the hysteresis between the touchdown and take-off conditions. Figure 2(d) shows the hysteresis as a function of dithering amplitude. For a given oscillation frequency, the low hysteresis is achieved only when exceeding a threshold value. Below the threshold value, the hysteresis reduction is not exhibited. It is worth mentioning that although the interface voltage increases the attractive Coulombic force, the AC modulation induced head oscillation is able to completely suppress the hysteresis.     

To understand the impact of imposed tunable oscillation through electrostatic interaction on the contact hysteresis, we consider a simple lumped parameter model in the framework of hysteresis during contact. Note that the head-disk interface has six orders of magnitude higher sliding speed than AFM based studies and does not suffer from the stick-slip phenomenon. Contacts in the head-disk interface is very unique as the events happen at very high sliding speed leading to a vibrational response of the head. The frequency of this vibration is driven by the contact stiffness. The resulting force on the head lags a certain amount of time before its effect can be felt. The force acting at the bulge of the deformable slider, takes some time for the induced wave to propagate within the slider.  In the model, we account for the forces acting at the head-disk interface: inter-molecular force, electrostatic force, air-bearing force, and the elastic contact between two bodies. The air bearing force is an uplifting force required to keep the head afloat on top of rotating disk. Further, the elastic contact force arises due to the interaction between the head and the rough surface on the disk. Contacts occur near the peaks of the roughness on the disk, so the distribution of peaks is the relevant variable, not the actual roughness of the disk. The peak distribution is assumed to follow a Rice distribution. A further refinement to the contact model is provided by observing that the contact area of the head is limited to the thermal actuator tip, about 10 $\mu m$ wide, and thus the interaction with the disk is only with a few peaks.  

The voltage induced electrostatic force increases the attractive electrostatic force between the head-disk interface given by:
\begin{equation}
F_{el}=\frac{\epsilon V^{2}}{2}\int_{d}^{\infty} \frac{P(z)}{(d-z)^{2}} dz
\end{equation}
where $P(z)$ is the spacing probability distribution, $\epsilon$ is the dielectric permittivity, $d$ is the initial clearance between the head-disk interface, and  $V$ is the applied voltage: DC component $V_{dc}$ and AC component $V_{AC}$ such that $V=V_{DC}+V_{AC}sin(wt)$. The electrostatic force is an attractive force which reduces the fly height (parabolic dependence) until the pull-in threshold is reached \cite{RajauriaSR16, KniggeIEEE04}. At this point the effective stiffness becomes singular, and the slider thermal actuator makes contact with the disk. Figure 2(b) shows the change in the estimated initial fly height as a function of DC voltage at the head-disk interface along with the parabolic fit. The estimated clearance change is measured by monitoring the change in heater power required to make a contact between the head and disk. The change in initial flying clearance is symmetric to zero voltage (as expected from Equation 1).  Note that the touchdown/take-off cycle with DC voltage does not remove hysteresis.

%******************************************************************************************************
\begin{figure}[htbp]
\begin{center}
\includegraphics[width=3.6in]{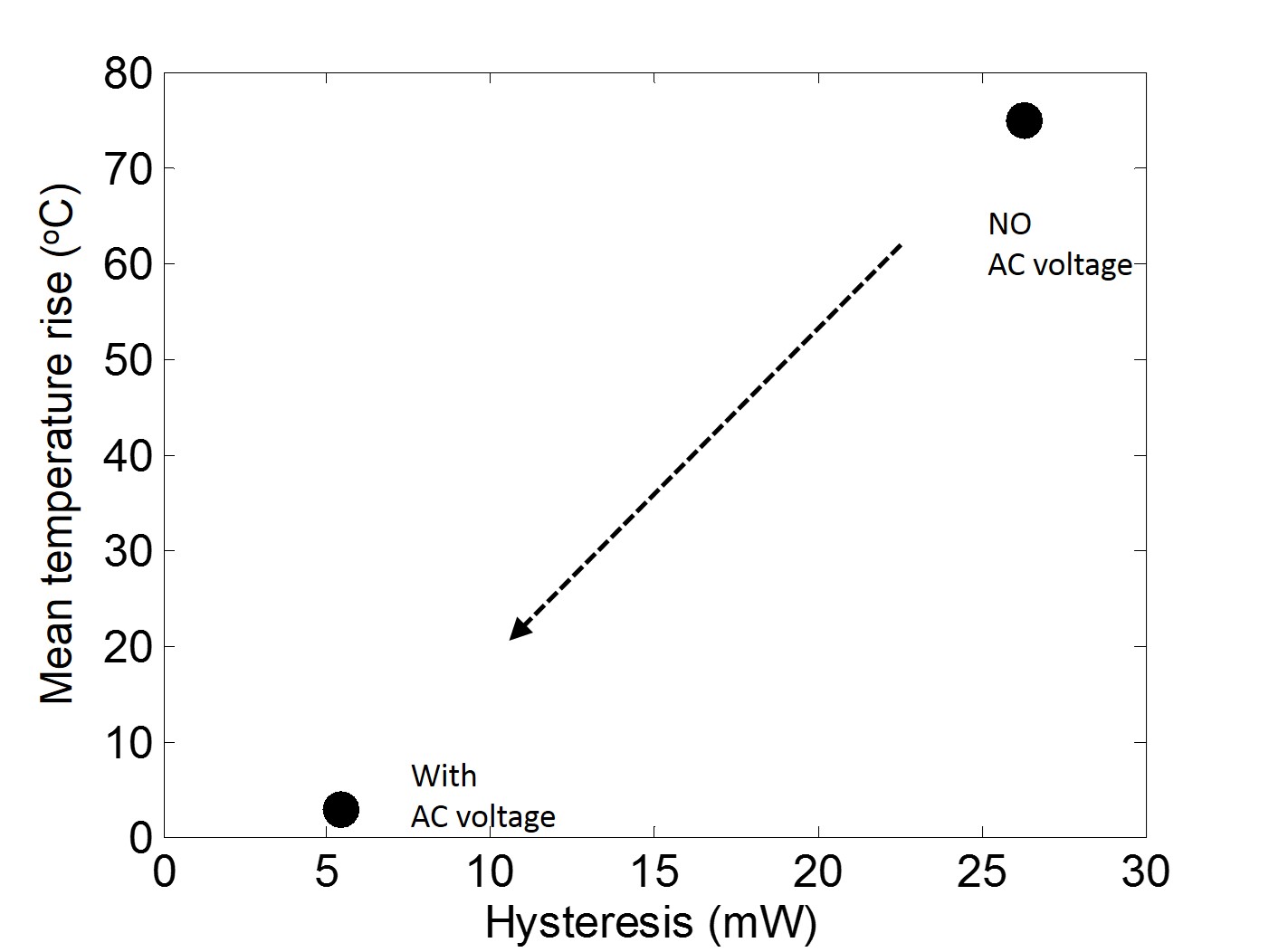}
\end{center}
\caption {Shows the increase in mean temperature rise as a function of hysteresis.}
    \label{fig:4}
\end{figure}
%******************************************************************************************************

The intermolecular force at the interface has an attractive $H_{a}$ and a repulsive component $H_{r}$, with $H$ being the Hamaker constant. It can be written as:
\begin{equation}
F_{m}=\int_{d}^{\infty} \Bigg(\frac{H_{a}}{6\pi}\frac{1}{(d-z)^{3}}-\frac{H_{r}}{45\pi}\frac{1}{(d-z)^{9}}\Bigg) P(z) dz
\end{equation}
Note that both electrostatic and intermolecular forces act when there is a separation in between the head and disk. For the contact between the head and the disk, the contact force can be written as:
\begin{equation}
F_{c}=\frac{4EN_{o}R^{1/2}}{3(1-\nu)^{2}}\int_{d}^{\infty} (z-d)^{3/2}f_{p}(z) dz
\end{equation}
where $E$ is the elastic modulus, $N_{o}$ is the asperity density which is assumed to be 150/$\mu m^{2}$, $R$ is the asperity ratio, $f_{p}(z)$ is the peak probability distribution, and $\nu$ is the Poisson ratio. It is worth noticing at low clearance and for low roughness standard deviation the molecular force is dominant, whereas for high roughness the contact force dominates. The electrostatic force follows the molecular force trend. The disk roughness is critical to the hysteresis operating regime. 

With this model, the rigid body equation of motion of the head is solved at every clearance to extract the state of the head. Figure 3 shows the simulation of the AC voltage impact on contact hysteresis with parameters similar to the experiment. Figure 3(a) and Figure 3(c)  show the complete time domain force between the head and disk at 0 V and 0.6 V AC volts, respectively. To emulate the experiment, a micro-heater embedded in the head is used to make a contact in between the head and disk. Such contact is made at t$\sim$0.5 $ms$ leading to a large variation in the resultant force. As the head to disk spacing decreases (red line), both the electrostatic (red line) and intermolecular forces become negative, pulling the head towards the disk (shown by Figure3(b)). When they touch, the contact force increases sharply (purple line). Further, at t$\sim$2.5 $ms$ the micro-heater power is reduced to pull the head away from the disk. This power reduction is the measure of contact hysteresis. While in contact the head response is similar with and without AC voltage. However, the effect of AC voltage is evident as the micro-heater power is reduced (t$\sim$2.5 $ms$) to retract the head from the disk. Figure 3(d) shows the hysteresis as a function of the AC voltage. Hysteresis is high at low AC voltages and decreases sigificantly with increased voltage amplitude. The agreement is very good between the simulation and the experiment (Figure 2(d)), the reduction in hysteresis is a function of  AC voltage bias. Let us emphasize that the agreement applies to the unique system of the head-disk interface inside the commercial hard disk drive where the sliding speed is six orders of magnitude higher than in AFM based studies.         

This remarkable agreement shows that although the electrostatic force (DC or AC voltage) is attractive, the AC voltage leads to an out-of-plane sub-nanometer oscillation of the head respect to the disk. This induced oscillation is able to completely suppress the contact hysteresis. To further quantify the impact of AC oscillation, we also measured the rise in head temperature during contact. Inside the head is an embedded contact sensor (ECS) to measure the localized rise in temperature during contact  with the disk caused by frictional heating \cite{XuIEEE14}. The temperature increase ($\Delta T_{fric}$) can be estimated by the ratio of frictional power dissipated into the sliding interface to the rate at which this heat is conducted away. It is given by: $\Delta T_{fric}=\frac{F_{fric}.v}{4ak}$, where $F_{fric}$ is the frictional force, $a$ the radius of contact, $k$ the thermal conductivity and $v$ the disk velocity. Figure 4 shows the rise in head temperature as a function of contact hysteresis. The reduction in contact hysteresis translates into a siginificant reduction in average head temperature. This futher confirms the importance of out-of-place oscillations in suppressing contact hysteresis at high sliding speeds.

In conclusion, we have performed a quantitative analysis of high speed contact events at the head-disk interface of commercial hard disk drives. It demonstrates that although the electrostatic force (DC or AC voltage) is attractive in nature, the low frequency AC voltage induced excitation is able to effectively remove hysteresis. It proposes a method to attain hysteresis-free contacts and reduced wear. This is of practical importance and of great economic value to the industry as well.

Authors acknowledge R. Smith for careful reading of manuscript.

 \end{document}